\documentclass[12pt]{article}
\usepackage{graphicx}
\usepackage{graphics}
\newcommand{\beq}{\begin{equation}}
\newcommand{\eeq}{\end{equation}}
\newcommand{\beqs}{\begin{eqnarray}}
\newcommand{\eeqs}{\end{eqnarray}}

\newcommand{\indi}{\nu_1\cdots\nu_{p+1}}

\newcommand{\indii}{\nu_1\cdots\nu_{p+2}}
\begin{document}
\begin{titlepage}
\begin{flushleft}
%       \hfill                      {\tt hep-th/000****}\\
       \hfill                       FIT HE 8-01/002 \\
\end{flushleft}
\vspace*{3mm}
\begin{center}
{\LARGE Beta-functions in Yang-Mills Theory \\ } 
\vspace*{5mm}
{\LARGE   from Non-critical String \\}  
\vspace*{12mm}
{\large Kazuo Ghoroku\footnote{\tt gouroku@dontaku.fit.ac.jp}}\\
\vspace*{2mm}

\vspace*{2mm}

\vspace*{4mm}
%{\it\large Fukuoka Institute of Technology, Wajiro, Higashi-ku}\\
%{\it\large Fukuoka 811-0295, Japan\\}
{\large Fukuoka Institute of Technology, Wajiro, Higashi-ku}\\
{\large Fukuoka 811-0295, Japan\\}
\vspace*{10mm}
\end{center}

\begin{abstract}

The renormalization group equations of the
Yang-Mills theory are examined in the non-critical string theory
according to the framework of the holography.
Under a simple ansatz for the tachyon,
we could find several interesting solutions which are classified by the value
of the tachyon potential at the vacuum. We show 
two typical, asymptotic-free solutions which are different in their
infrared behaviors. For both types of solutions, we 
could obtain quark-confinement
potential from the Wilson-loop. The stability of the tachyon and the
ZigZag symmetry are also discussed for these solutions.

\end{abstract}
\end{titlepage}

\section{Introduction}

~~The idea of string description of the super-symmetric gauge theory
has been proposed firstly by Maldacena \cite{M1}
and developed to the case of
non-supersymmetric case by several ways ~\cite{GKP1,W1,W2}. 
One step in this direction has been proposed by \cite{Poly1,GKP1} 
in terms of the non-critical string theory, which is based on
the super-symmetric Liouville theory, and developed ~\cite{FM,FM2,AFS,AG,G1}.
This proposal has studied also in the critical type 0B
~\cite{KT,KT3,KT4,Min,Min2,GLMR} and also in the type 
IIB ~\cite{KS1,NO,Gub,GPPZ,CM} string theories.

In the approaches based on the non-supersymmetric global theory, 
several asymptotic solutions have been discussed
both in the ultraviolet and infrared regions. And
many authors have tried to connect 
these fixed points by one renormalization group flow. In order to
clear this point, we should find such solutions,
which connect those fixed points,
by solving the renormalization group equations. Our purpose here is to show
such a solution explicitly in the framework of the non-critical string
theory for the pure Yang-Mills theory, where ZigZag invariance would be
expected \cite{Poly1,AG}.

The solutions obtained here are classified 
into the asymptotically free and non-free
types. The asymptotic-free solutions are further separated into
two types by their infrared behaviors. One type of solutions
has an infrared fixed point with the anti-de Sitter
(AdS) background at a finite coupling constant, where the ZigZag invariance
is realized. 
While the $\beta$-function of the other type decreases monotonically
with the increasing coupling-constant.
Estimating the Wilson loops for these asymptotic free
solutions, we could obtain the quark-confining potential. 
The solutions of the $\beta$-function for non-zero tachyon potential
show unexpected behaviors near the asymptotic free region. We discuss
on this point from the viewpoint of the gauge theory.

In section two we give the gravitational equations to be solved as the 
renormalization group equations of the Yang-Mills theory. And 
the conformal invariant solution with AdS background is shown. The equation
to be solved is given as the one of $\beta$-function in section three, and
the running solutions with and without this AdS fixed point are obtained in 
the next two sections by assuming that the tachyon is independent on the
energy-scale. In the section six, the $\beta$-functions are discussed
in the series expanded form. 
In section seven, the stability of the tachyon is discussed, and
the concluding remarks  are given in the final section.

\section{Gravitational equations and conformal fixed point}

~~The effective action of the non-critical string theory, which is dual to
the Yang-Mills theory, could be represented by including the 
Ramond-Ramond (RR) $p+1$-form field $A_{p+1}$.
\footnote{ For the sake of the simplicity, we consider only one type of R-R 
field.
} 
And it is expected that N $D_{p+1}$-branes are stacked on the boundary to make
the U(N) gauge theory there. 
Then we start from the following action,
\beqs
    S_D &=& {1\over 2\kappa^2}
      \int dx^D\sqrt{|g|}\Bigg\{e^{-2\Phi}\left(R-
      4(\nabla\Phi)^2+(\nabla T)^2+V(T)+c \right) \nonumber \\ 
    &{}&  \qquad\qquad~~~~~~ + \frac{1}{2(p+2)!}f(T)F_{p+2}^2\Bigg\},
    \label{action} 
\eeqs
where $c=-(10-D)/2\alpha'$, and $F_{p+2}=dA_{p+1}$ is the 
field strength of $A_{p+1}$. Hereafter we take as $\alpha'=1$. 
The total dimension
$D$ includes the Liouville direction, which is denoted by $r$.
The tachyon potential is represented by
$V(T)$, and $f(T)$ denotes the couplings between the tachyon and the RR
field investigated in \cite{KT}. Although they might be
important, we don't know
their precise form. So we search for solutions in the case of $f(T)=1$ 
and $T=T_0=$ constant, for the sake of the simplicity.
And the value of $V(T_0)$ is taken as a parameter, which plays an
important role to determine the behaviour of the renormalization group flows
of the solutions.

The equations of motion are written as
\beqs
     R_{\mu\nu} - 2 \nabla_\mu \nabla_\nu \Phi &=& -\nabla_\mu T\nabla_\nu T + 
             e^{2\Phi} T_{\mu\nu}^A \label{metric}\\
    4 \nabla_\mu \Phi \nabla^\mu \Phi 
       -2 \nabla^2 \Phi &=& \frac{D-2d-2}{4(p+2)!}e^{2\Phi}F_{p+2}^2
            + V_c(T) \label{dilaton}\\
    \nabla^2 T -2 \nabla_\mu \Phi \nabla^\mu T 
         &=& \frac{1}{2}V'_c(T) \,
            \label{tachyon}
\eeqs
\beq
     \partial_\mu (\sqrt{|g|}F^{\mu\indi}) = 0 \, , \label{form}
\eeq
where $V_c(T)= c+V(T)$ and
\beq  
 T_{\mu\nu}^A = -\frac{1}{2(p+1)!}
                \left(F_{\mu\indi}F_\nu^{\;\;\;\;\indi}
       - \frac{g_{\mu\nu}}{2(p+2)}  F_{\indii}F^{\indii}\right).
     \label{tensor3} 
\eeq
For the simplicity,
the dimension $D$ is set as $D=p+2$ and $p=3$ to consider the case dual to
the pure 4-dimensional Yang-Mills theory.

We solve the above equations according to the following ansatz;
\beq
 ds^2= e^{2A(r)}\eta_{\mu\nu}dx^{\mu}dx^{\nu}
           +e^{2B(r)}dr^2 \, \label{metrica}
\eeq
\beq
    \Phi\equiv\Phi(r), \quad \quad T\equiv T(r) \quad \hbox{and} \quad
     A_{01\cdots p} = -e^{c(r)}\ , \label{RR}
\eeq
where $x^{\mu}, \mu=0\sim p$, denote the space-time coordinates.
And $r$ represents the Liouville coordinate, which
plays the role of the energy-scale in the $p+1$-dimensional gauge theory
\cite{Poly1}.
The equation (\ref{form}) is solved as
\beq
 \partial_{r}e^{c(r)}= N e^{dA+B} \, \label{RRfi}
\eeq
where $d=p+1$ and N denotes the number of the p-brane.
Then the remaining equations (\ref{metric}) and (\ref{dilaton}) are 
rewritten as,
\beq
    -\dot{A}\dot{B}+\ddot{A}+d\dot{A}^{2}-2\dot{A}\dot{\Phi}=
       {N^2\over 4} e^{2B+2\Phi}, \label{metric1}\\
\eeq
\beq
       d(\ddot{A}+\dot{A}^2-\dot{A}\dot{B})
          -2(\ddot{\Phi}-\dot{B}\dot{\Phi})=
       {N^2\over 4} e^{2B+2\Phi}-\dot{T}^2,  \label{metric2}\\
\eeq    
\beq
         2(\ddot{\Phi}+\dot{\Phi}(d\dot{A}-\dot{B}))-4\dot{\Phi}^2
      = {(d+1)\over 4}N^2 e^{2B+2\Phi}+e^{2B}V_c(T) \,
                 \label{dilaton2}\\
\eeq
\beq
    \ddot{T}+(d\dot{A}-\dot{B})\dot{T}-2\dot{\Phi}\dot{T}=
        {1\over 2}e^{2B}V'_c(T) \,
         \label{tachyon2}
\eeq
where the dot denotes the derivative with respect to $r$.
%%%%%%%%%%%%%%%%%%%%%%%%%%%%%%%%%%

We notice that the above equations have the AdS 
as a conformal invariant fixed point. 
We review it briefly. The solution is found
by assuming that $\Phi(=\Phi_0)$ and $T(=T_0)$ are independent on $r$
and 
\beq
  A= \gamma\rho
       , \ \quad B=-\rho \, , \label{ansatz}
\eeq
where $\rho=\ln(r/r_0)$ and $r_0$ denotes an appropriate scale parameter 
which is taken as unit hereafter for the simplicity.
Then we get
\beq
   \lambda_0\equiv Ne^{\Phi_0}=\sqrt{-{4\over 5}V_c(T_0)} \ , 
      \quad  \gamma= \pm{\lambda_0\over 4} \, ,
      \quad V'(T_0)=0. \label{dil}
\eeq
where $\lambda_0$ denotes the 't Hooft coupling constant at this fixed point. 

\section{Equations for $\beta$-Function}

Usual way to find renormalization group flows is to see the deviations
from the above AdS solution by taking the following functional forms,
\beq
  A(\rho)=\gamma\rho+a(\rho), \qquad B(\rho)=-\rho+b(\rho), \label{an1}
\eeq
\beq
  \Phi(\rho)=\Phi_0+\phi(\rho), \qquad T(\rho)=T_0+t(\rho), \label{an2}
\eeq
where $\gamma, \Phi_0$ and $T_0$ are given in 
(\ref{dil}). This setting is useful for
finding the fixed point with the AdS background, where
the deviations, $ \{a(\rho), b(\rho), \phi(\rho), t(\rho)\}$, vanish at
$\rho=\pm\infty$. 

While we want to search for more general solutions which are not 
necessarily asymptotic
AdS background. Then we use $A(\rho)$ instead of $a(\rho)$ here.
The equation for $T$, (\ref{tachyon2}), is solved as $t(\rho)=0$ 
since we are considering the case of $T=T_0$. Then
the remaining equations (\ref{metric1}) $\sim$ (\ref{dilaton2}) 
are rewritten by using $ \{A(\rho), b(\rho), \phi(\rho)\}$ as follows:
\beq
      \ddot{A}+\dot{A}(d\dot{A}-\dot{b}-2\dot{\phi})
  ={\lambda_0^2\over 4}
       e^{2(b+\phi)}, \label{metric11}
\eeq
\beq
       d[\ddot{A}+\dot{A}(\dot{A}-\dot{b})]
           -2(\ddot{\phi}-\dot{b}\dot{\phi})
             ={\lambda_0^2\over 4} e^{2(b+\phi)}, 
                \label{metric21}
\eeq    
\beq
    \ddot{\phi}+\dot{\phi}(d\dot{A}-\dot{b}-2\dot{\phi})
     = {5\lambda_0^2\over 8}
       e^{2(b+\phi)} +{V_c(T)\over 2}e^{2b} \, ,
                 \label{dilaton21}
\eeq
where the dot denotes the derivative with respect to $\rho$.

By introducing the following new notations
\beq
  \dot{A}=Q , \qquad \dot{\phi}=S,   \label{qs}
\eeq
these three equations can be rewritten into the two first order
differential equations and one constraint as follows
\beqs
     \dot{Q}-Q\dot{b} &=& b_1  , \label{diff11}\\
     \dot{S}-S\dot{b} &=& b_3 \, ,\label{diff12}
\eeqs
\beq
   b_2-db_1+2b_3=0 ,  \label{constr}
\eeq
where
\beqs
     b_1 &=& {\lambda_0^2\over 4} e^{2(b+\phi)}-Q(dQ-2S) \label{eqQ}\\
     b_2 &=& {\lambda_0^2\over 4} e^{2(b+\phi)}-dQ^2     \label{eqS}\\
     b_3 &=& {5\lambda_0^2\over 8} e^{2(b+\phi)}-S(dQ-2S) +
                \frac{1}{2}V_c(T_0)e^{2b} \,  .          \label{eqb}
\eeqs
We can see that the equation (\ref{constr})
represents the "Hamiltonian" constraint
which comes from the reparametrization invariance of the gravitational 
system with respect to
the coordinate $r$ or $\rho$ .
In fact, the action (\ref{action}) can be written as the kinetic term ($K$)
minus the potential term $U$ by using the ansatz (\ref{metrica}) $\sim$
(\ref{RRfi}) as
\beqs
  K&=&\sqrt{g}\left(4(3\dot{A}^2-4\dot{A}\dot{\Phi}+\dot{\Phi}^2)
          e^{-2(b+\Phi)}+N^2/2 \right), \label{kinetic} \\
  U&=&\sqrt{g}V_c(T_0)e^{-2\Phi}, \label{pote}
\eeqs
where we used $T=T_0$=const.. Then the zero-energy constraint, $T+U=0$,
is written as
\beq
  4(3\dot{A}^2-4\dot{A}\dot{\Phi}+\dot{\Phi}^2)+({\lambda^2\over 2}+
     V_c(T_0))e^{2b} = 0, \label{constr2}
\eeq
where  $\lambda=Ne^{\Phi}$, which is the running 't Hooft coupling constant.
This equation is equivalent to
(\ref{constr}). Usually the gauge $b=0$ is considered,
and (\ref{constr}) is used
as the boundary condition to solve the equations of $A$ and $\Phi$.

Here, $b$ is not however fixed for a while. And
the equation (\ref{constr2}) (or (\ref{constr})) is used to eliminate $b$ 
in the equations (\ref{diff11}) and (\ref{diff12}). After that,
we find that (\ref{diff11}) and (\ref{diff12}) are
not independent, and we obtain a trivial identity ($0=0$)
and the following equation
\beq
  (1-Sb_{1S})\dot{S}-Sb_{1Q}\dot{Q}=b_3+Sb_{11} \, , \label{diff3}
\eeq
where
\beqs
     b_{1Q} &=& {3Q-2S\over 3Q^2-4QS+S^2} \label{b1Q}\\
     b_{1S} &=& {-2Q+S\over 3Q^2-4QS+S^2} \label{b1S}\\
     b_{11} &=& -{S\lambda^2\over 2V_c+\lambda^2}.  \label{b11}
\eeqs
This implies the following facts. 

(i) One of the two functions,
$Q$ or $S$, can be given as an arbitrary function
and the remaining one should be obtained by solving the
equation (\ref{diff3}). This arbitrariness would come from
the fact that $b$ is not yet fixed at this stage.

(ii) As far as the condition (\ref{constr2}) is satisfied,
it is enough to solve the equation (\ref{diff3}) for obtaining
the solutions of the original equations.

Our strategy is to rewrite Eq.(\ref{diff3}) as the equation of
$\beta$-function by fixing $b$ and determining $Q$ in an appropriate way
as shown below. Before fixing $b$, we define
the $\beta$-function of the Yang-Mills theory as follows
\beq
       \beta(\lambda)\equiv \dot{\lambda} ,
           \qquad .  \label{beta}
\eeq
The functions to be solved ($Q$ and $S$) are originally
introduced as functions of $\rho$, and
the 't Hooft coupling is also the function of $\rho$ as 
$\lambda=\lambda(\rho)$.
So it would be possible to consider 
$Q$ and $S$ as functions of $\lambda$ if 
$\lambda$ could be regarded as a single valued function of $\rho$.
Then the variable $\rho$ and the function $S$ in the equation (\ref{diff3}) 
can be replaced to $\lambda$ and $\beta$ by using the relations,
$d/d\rho=\beta(\lambda)d/d\lambda$ and
$S=\dot{\Phi}=\beta/ \lambda$. 

Next we determine $Q$ as a function of $\beta$ and $\lambda$ 
by fixing $b$ as follows. Consider a new coordinate $u$ defined as
\beq
  e^{2b}d\rho^2=du^2 . \label{newv}
\eeq
For this coordinate,
we can define new $\beta$-function as $\tilde{\beta}=d\lambda/du$.
We restrict the analysis to the case where both functions $\beta$ and
$\tilde{\beta}$ represent essentially the same $\beta$-function.
Then we assume $\beta/\tilde{\beta}=1$ simply. This assumption leads to the 
gauge, 
\beq
  ({du \over d\rho})^2=e^{2b}=1. \label{cons2}
\eeq
From this we obtain, 
\beq
    Q={1\over 3\lambda}\left(2\beta\pm
         \sqrt{\beta^2-{3\over 8}\lambda^2(\lambda^2+2V_C)}\right). \label{QQ}
\eeq
Using this, Eqs.(\ref{diff3}) is rewritten as
\beq
 \beta'={\beta\over 3\lambda}+{4V_c+5\lambda^2\over 8\beta}\lambda
               \mp {4\over 3\lambda}\sqrt{\beta^2-{3\over 8}\lambda^2
                    (\lambda^2+2V_c)},
         \label{betafu2}
\eeq
%%%%
where prime denotes the derivative with respect to $\lambda$, for example
$\beta'=d\beta/d\lambda$. $V_c=V_c(T_0)$ is a constant. 
%%%%
The sign $\mp$ of the third term
on the right hand side implies that if $\beta$ would be 
a solution of the equation of either sign then $-\beta$ is the
solution of the equation of the opposite sign. In this sense, it is enough
to consider the equation of either sign. 

In deriving equation (\ref{betafu2}), the overall factor $(3Q-2S)$ has been
divided out since we can see that $3Q=2S$ is not the solution of the
original equations.

\section{Analytic solutions: $V_c(T_0)=0$}

First, we consider the analytic solution obtained
for $V_c=0$. It is easy to find the following solution, 
\beq
 \beta = {5\over 2}\lambda Q=-3\sqrt{25\over 24}\lambda^2 ,   \label{sol03}
\eeq
where we solved the equation (\ref{betafu2}) with the minus sign of
the third term. This is the exact solution, but it contains only 
one-loop term. In this sense, this corresponds to the $\beta$-function 
of the super-symmetric Yang-Mills theory, but this should be considered as
the approximate form at small $\lambda$ in our model as mentioned below.

For this solution, we obtain $e^A=\lambda^{2/5}$.
Then the background in the asymptotic free
limit of this solution is not the $AdS_5$ as found in ~\cite{Min2,KT3},
and it is considered as the "ZigZag" horizon ~\cite{Poly1,AG} since 
$e^A\to 0$ in the limit of $\lambda\to 0$. 
This horizon is situated at the ultraviolet limit, $\rho\to \infty$, contrary
to the expectation of \cite{AG}. As for
the five dimensional scalar curvature $R^{(5)}$, it
is obtained as $R^{(5)}=R_0\lambda^2$ with $R_0=20/3$.
Then $R^{(5)}$ grows with increasing $\lambda$, so the
higher-order terms of the curvature would become important
in the effective action at large $\lambda$.
And there is no reason to prevent those terms in our model. 
As a result this solution would
be modified in the infrared region 

%%%%%%%%%%%%%%%%%%%
This point is understood from the viewpoint of the availability of
the holographic method considered here.
For $D=10$ and $p=3$, the conditions of the reliability of correspondence
between the gravitational equations and the renormalization group 
equations of the
Yang-Mills theory are given by $N, \lambda >>1$. While they are 
changed to $N>> \lambda^{5/2}$ and $1 >>\lambda$ if we take the setting,
$D=p+2$ and $p=3$, seriously. The latter condition, $1 >>\lambda$, is
obtained from the requirement that the higher order terms of $\alpha'$
in the effective action are negligible in the D-brane system. It can
be expressed as $\lambda^{1/(D-p-3)}>>1$ for general $D$ and $p$.
Then the higher curvature terms would be necessary to apply the holographic
picture at large $\lambda$ region contrary to the case of the critical
string theory.

%%%%%  inserted A %%%%%%%%%%%%%%
From the logarithmic behaviour of the gauge coupling constant in the
perturbative region, 
the result given by Eq.(\ref{sol03}) implies that the gauge coupling
$g_{\rm YM}$ is related to the 'tHooft coupling $\lambda$ as
\beq
  \lambda \propto g_{\rm YM}^2 .
\eeq
This relation is expected from the Born-Infeld type action of 
the D-brane, which implies $e^{\Phi}\propto g_{\rm YM}^2$. However
this relation is altered in the case of $V_c\neq 0$ as seen in the next
section, and we will discuss on this point in more detail in the
section six.
%%%%%%%%%%%%%%%%%%%%%%

Although the solution would be unreliable at large $\lambda$, we can estimate
the Wilson loop since the solution is exact.
In \cite{M2}, the Wilson-loop is expressed by the
integral with respect to the coordinate transverse. Here,
we rewrite this
expression by replacing the integral variable from the coordinate to the
coupling constant $\lambda$ as
\beq
   \int_{r_0}^{r_M}{dr\over r}=-\int_{\lambda_{M}}^{\lambda_{IR}}
              {d\lambda\over \beta(\lambda)}, \label{trans}
\eeq
where $\lambda_{IR}$ ($\lambda_{M}$) represents the upper (lower) bound
of $\lambda$ at the infrared (ultraviolet) side. Then
the distance $L$ between $q\bar{q}$ and the extremized
string action $S$ are given as
\beq
 L=-2Er_0\int_{\lambda_{M}}^{\lambda_1}{d\lambda\over \beta}
       {e^{-A+b}\over \sqrt{e^{4A}-E^2}}, \label{Le}
\eeq
\beq
 S=-{\tau r_0\over 2\pi}\int_{\lambda_{M}}^{\lambda_1}{d\lambda\over \beta}
       {e^{3A+b}\over \sqrt{e^{4A}-E^2}}, \label{Se}
\eeq
where $E$ is an arbitrary constant introduced in minimizing the string 
action \cite{M2}, and $\tau$ denotes the time-interval of the Wilson-loop.
Here the lower bound is given as $\lambda_{M}=E^{5/4}$ since 
$e^A=\lambda^{2/5}$, and the upper bound is infinite.
Then we obtain the following result,
\beq
    S = S_3 L^{3/7} \,, \quad S_3=\tau{\sqrt{3/2}r_0 B(3/8,1/2)\over 
                 4\pi[\sqrt{3/2}r_0 B(11/8,1/2)]^{1/5}}
                .   \label{potsup}
\eeq
This result shows the confinement potential, but $S$ increases more slowly 
with $L$ than the linearly rising potential which is expected from the
well known QCD. 
The second unexpected point is that the potential does not show the 
Coulomb behaviour
near $L=0$. This point could be understood as follows. We obtain 
$L=L_0E^{-7/4}$, where $L_0$ is a positive constant, then the 
potential at small $L$ is determined by the information at large $E$.
While the lower bound of $\lambda$ is given by $\lambda_{M}=E^{5/4}$, so
the behaviour
of the potential at small $L$ is given by the large coupling dynamics.
As a result the Coulomb behaviour could not be seen at small $L$.

Where is the AdS fixed point in this solution? This solution is obtained
for $V_c(T_0)=0$, then the
AdS fixed point at $\lambda_0 (=\sqrt{-4V_c/5})$ is pushed to
$\lambda=0$. And the scalar curvature becomes zero since the radius of the
AdS space becomes infinite.
In this sense, the usual AdS solution could not be seen. 
As shown in the next section, it is necessary to consider 
the case of $V_c<0$ in order to find a solution including the AdS 
fixed point at finite $\lambda$.

\section{Numerical solutions: $V_c(T_0)<0$}

In the case of $V_c(T_0)<0$, we could show 
solutions which contain the AdS fixed
point by solving Eq.(\ref{betafu2}) numerically.
%%%%%%%%%%%%%%%%%%%%

Before performing the numerical analysis, we notice some points implied by
Eq.(\ref{betafu2}) to consider an appropriate boundary
conditions to solve the equation. (i) Firstly, it can be read
that two fixed points are possible at $\lambda=0$ and $\lambda_0 (=
\sqrt{-4V_c/5})$. If $\beta$ would be zero at some other point,
then the third term of the right hand side of (\ref{betafu2}) diverges.
So the solutions for $\beta$ could not have zero points except for the
above two points.
(ii) Secondly, Eq.(\ref{betafu2}) 
represents two equations discriminated by the
sign of the third term. As mentioned above, this freedom can be reduced to
the sign of $\beta$. Then it is enough to solve the equation in either
sign since the other solution can be obtained by reflecting its sign.
(iii) Thirdly,
$\beta$ could be obtained near $\lambda=0$ as
\beq
 \beta=\beta_0^{\pm}\lambda+\cdots,  \qquad
        \beta_0^{\pm}=\pm{1\over 2}\sqrt{-V_c}.     \label{AFlim}
\eeq
The coefficient should be taken as 
$\beta_0^{-}=-{1\over 2}\sqrt{-V_c}$ for the 
asymptotic-free solution. But we notice that this result is a little 
undesirable since the leading 
term is not quadratic with respect
to $\lambda$ as expected from the perturbative Yang-Mills theory. 
The meaning of this term and the series expansion of $\beta(\lambda)$
are discussed in the next section.

(iv) Near the AdS fixed point, $\lambda=\lambda_0$, the following
expansion form can be found,
\beq
     \beta = \beta'_{\pm}(\lambda_0)(\lambda-\lambda_0)+\cdots,
          \qquad \beta'_{\pm}(\lambda_0)={-1\pm\sqrt{6}\over 2}\lambda_0.
                      \label{AdSlimit}
\eeq
This result indicate that the AdS fixed point is either the ultraviolet or
the infrared fixed point depending on the sign of $\beta'(\lambda_0)$.
In principle, both solutions are possible.

We solve Eq.(\ref{betafu2}) with the minus sign of the third term
of its right hand side since the asymptotic free solutions can be obtained
by this choice. The value of the potential is taken as $V_c(T_0)=-15/2$
for simplicity.
Although there are many solutions depending on the 
boundary condition, they are separated into two groups, the asymptotic free
and non-free solutions. 
The typical solutions are shown in the Fig.1. 
We should notice that the functions given
by the reflection, $\beta \to -\beta$, of the solutions shown in the
Fig.1 are also the solutions which are
not shown in the figure. 

%===  Fig.1  ========================================================
\begin{figure}
\begin{center}
   \includegraphics[width=15cm,height=7cm]{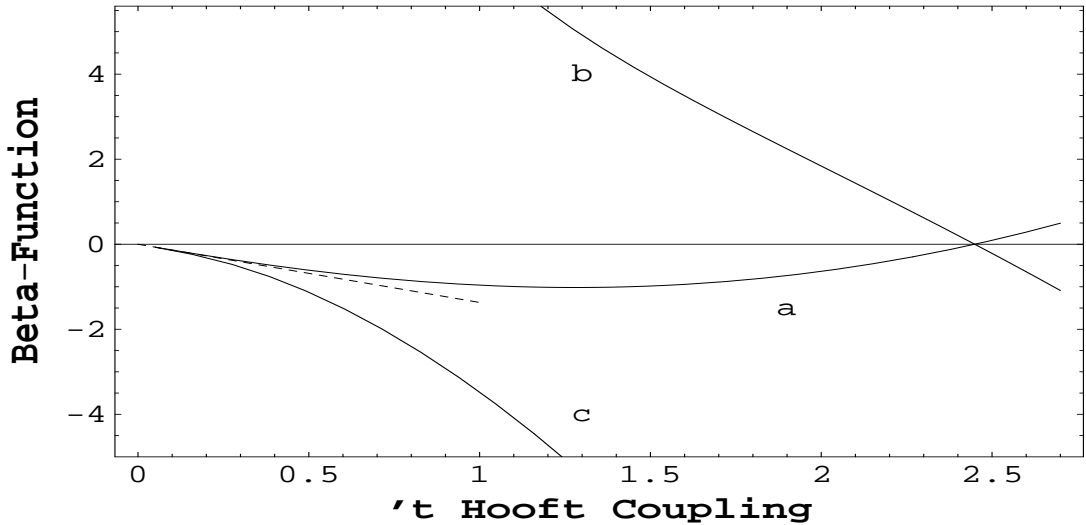}
 \caption{The numerical solution of $\beta(\lambda)$ obtained by 
(\ref{betafu2}) with the minus sign of the third term and 
$V_c=-15/2$. The three solid 
curve, (a), (b) and (c), are obtained by taking the boundary condition 
given in the text. The dotted line is the asymptotic solution approximated
near $\lambda=0$, $\beta=\beta_0^-\lambda$.}
\end{center}
\end{figure}
%====================================================================

The solution (b) has been obtained by the boundary condition,
$\beta(\lambda_0+\epsilon)=-\beta'_{+}(\lambda_0)\epsilon$, where 
$\epsilon$ is a very small number. This solution has the ultraviolet
fixed point at $\lambda_0$ and diverges to $\infty$ ($-\infty$) 
at $\lambda=0$ ($\lambda=\infty$). 

While the solution (a) is obtained
by the boundary condition,
$\beta(\lambda_0+\epsilon)=\beta'_{+}(\lambda_0)\epsilon$.
This condition is the reflection on the $\lambda$-axis of the above
boundary condition for the solution (b),
but it does not leads to the solution given by changing the sign of the
solution (b). The reflected solution of (b) is obtained from Eq.(\ref{betafu2})
of the plus sign of the third term with the same boundary condition for the
solution (b).

The solution (a) crosses the $\lambda$-axis at
$\lambda=\lambda_0$ as the infrared
fixed point and approaches to zero as $\lambda\to 0$ according to
the asymptotic solution
$\beta=\beta_0^{-}\lambda=-{1\over 2}\sqrt{-V_c}\lambda$ which is shown by the 
dotted line in the Fig.1. Then the asymptotic-free and the AdS infrared 
fixed points are smoothly connected by this solution.
We can further
see $e^{A(\lambda)}\to 1$ as $\lambda\to 0$, then the background at the 
asymptotic-free fixed point is not the AdS space but 
the flat space-time. So this solution denotes the renormalization group
flow between two different background configurations, flat and AdS spaces.
It should be noticed for this solution that $e^{2A}\to 0$ as
$\lambda\to \lambda_0$, near the AdS fixed point. This is seen from the
approximate form of $e^{2A}$ obtained near $\lambda\sim\lambda_0$,
\beq
  e^{2A}=e^{{1\over 2}\lambda_0\rho}
          =(\lambda_0-\lambda)^{\sqrt{6}+1 \over 5}.  \label{adsIR}
\eeq
Here we notice the following relation, 
$\rho=\ln (\lambda-\lambda_0)/\beta'_+(\lambda_0)$, in this region.
Then this fixed point can be considered as the ZigZag horizon, which is
situated at the infrared boundary $\rho=-\infty$ as expected in \cite{AG}.
The five dimensional scalar curvature $R^{(5)}$ is finite since this
point corresponds to the AdS space, and it is given by
$R^{(5)}={5\over 4}\lambda_0^2=|V_c|$ at $\lambda=\lambda_0$.

The third solution (c) in the Fig.1 is obtained by the condition, 
$\beta(\epsilon)=\beta_0^{-}\epsilon$, but it can not arrive at the AdS
fixed point and diverges to $-\infty$ at large $\lambda$. 
This solution has the similar form to the one
given in the previous section for $V_c=0$, but the asymptotic
form near $\lambda=0$ is different. As in the solution (a),
$(Q=) \dot{A}$ approaches to zero as $\lambda\to 0$ and we obtain $e^{2A}=1$
at $\lambda=0$, while $e^{2A}$ increases monotonically with increasing 
$\lambda$. Then the background of the asymptotic-free fixed
point of this solution is also the flat five dimensional space-time, and
there is no ZigZag horizon in this solution. 
These points are also different from the solution of $V_c=0$.

Next task is to evaluate the Wilson loop for these solutions.
For solutions (b) and (c), we need the information of the infinite
range of $\lambda$ for the analysis in the asymptotic-free phase.
But it seems to be difficult to do it since we do not know the 
analytic form of the solutions contrary to the case of the previous section.
While, the asymptotic free phase for the solution (a) is restricted to
the finite region of $\lambda$, $0< \lambda < \lambda_0$.
Then it is possible to estimate numerically
the Wilson loop for this solution according to the formula given above.
We notice some remarks in performing this numerical analysis.

(i) The form of $e^{A}$ obtained here is shown in the Fig.2.
It implies that the analysis should be performed in the region
$0<E<1$, and the upper limit in the integration of $\lambda$ should be
introduced for each $E$. 
%===  Fig.2  ========================================================
\begin{figure}
\begin{center}
   \includegraphics[width=13cm,height=7cm]{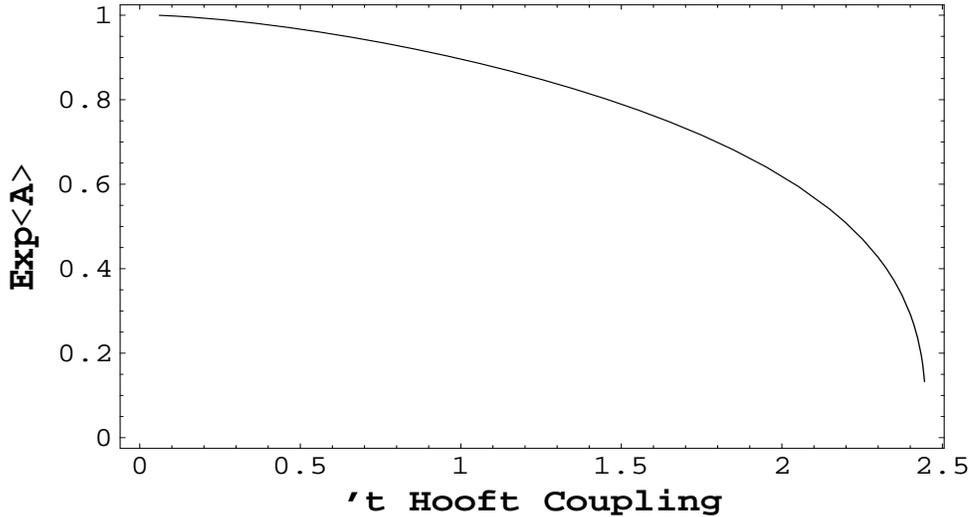}
 \caption{The form of $e^{A(\lambda)}$ for the solution (a)
is shown by the curve. The horizontal line is the $\lambda$-axis.}
\end{center}
\end{figure}
%====================================================================

(ii) $e^{A}$ approaches one at $\lambda=0$ as stated above, so the
integration has the logarithmic divergence near
$\lambda=0$ since $\beta\to \beta_0^{-}\lambda$ for $\lambda\to 0$. Then
we should evaluate the integrals by subtracting this divergence, and the
following subtracted form for $S$ and $L$ are used,
\beq
 L=-2Er_0\int_{\lambda_{M}}^{\lambda_1}{d\lambda\over \beta}
           (1-{\beta\over \beta_0\lambda})
       {e^{-A+b}\over \sqrt{e^{4A}-E^2}}, \label{Lesub}
\eeq
\beq
 S=-{\tau r_0\over 2\pi}\int_{\lambda_{M}}^{\lambda_1}{d\lambda\over \beta}
           (1-{\beta\over \beta_0\lambda})
       {e^{3A+b}\over \sqrt{e^{4A}-E^2}}. \label{Sesub}
\eeq

%===  Fig.3  beta1.nb (Fig.2B)   beta2.nb (Fig.B)   =========================
\begin{figure}
\begin{center}
   \includegraphics[width=14cm,height=7cm]{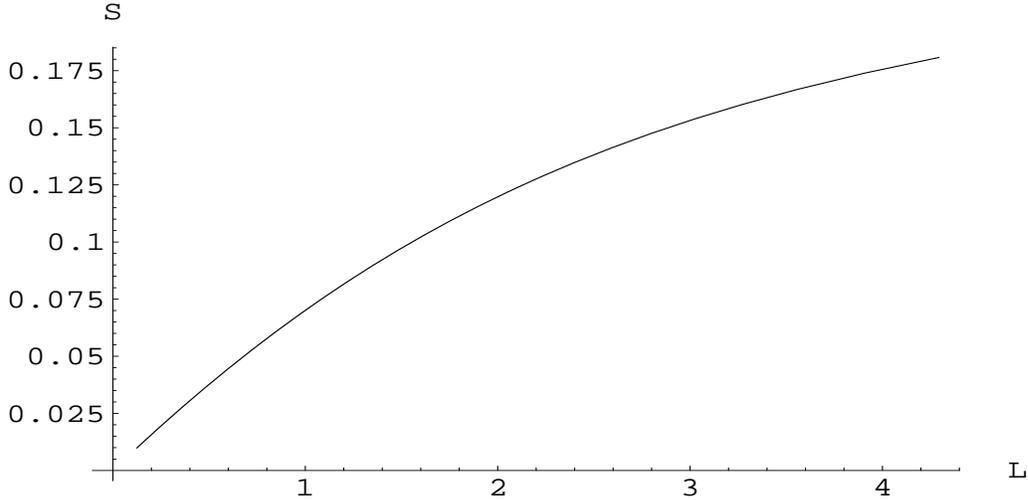}
 \caption{The numerical result of $Q\bar{Q}$
potential extracted from the Wilson-loop is shown for the solution
(a). The potential near $L=0$ is estimated analytically in the text.}
\end{center}
\end{figure}
%====================================================================

The result is shown in the Fig.3. The estimation in the region of small 
$L$ is suppressed
in this figure since a special care is needed in this region where
both $\beta$ and $e^{A}$ approach to zero. So we estimate
$S$ and $L$ analytically in this region as shown below. 
For $\lambda\sim\lambda_0$, we obtain
\beq
  \beta=-\beta'_{+}(\lambda_0)x, \qquad e^{A}=x^\alpha,
\eeq
where $x=\lambda_0-\lambda$, $\alpha=\lambda_0/(4\beta'_{+}(\lambda_0))$ and
the higher order terms of $x$ are suppressed. By using these approximate
formula, we obtain the following result
\beq
  S=-{B^2(3/4,1/2)\over 16\pi}{1\over L} \, .  \label{coulm}
\eeq
where $B(a,b)$ denotes the beta function. Then the attractive Coulomb potential
can be seen for this solution at small $L$, but this Coulomb behaviour
does not reflect the dynamics of the asymptotic free region. 
It should be the 
reflection of the dynamics near the infrared fixed point.
%%%%%%%%

\section{Behaviour of $\beta(\lambda)$ near $\lambda=0$}

Here we discuss the behaviour of $\beta(\lambda)$ near $\lambda=0$.
It is sharply discriminated by the value of $V_c$. For $V_c=0$,
the expected asymptotic-free behaviour and the favorable correspondence
between $\lambda$ and $g_{\rm YM}$ are obtained in the section four. 

However the situation
is very different in the case of $V_c < 0$. First, we comment on the
leading linear term of $\beta$ given in Eq.(\ref{AFlim}). From the
viewpoint of the field theory, the coefficient of this linear term 
represents the anomalous dimension of $\lambda$. This could be understood
when we apply the idea given in \cite{W1} to the background considered
here. From Eq.(\ref{action}), the following wave equation for $e^{-\Phi}$
is obtained,
\beq
  (-\nabla^2+m^2)\chi=0, \qquad m^2=-{1\over 4}(V_c+R), \label{anom}
\eeq
where $\chi=e^{-\Phi}$. $\nabla$ and the scalar curvature $R$ are given
by using the solution given in the previous section as follows,
\beq
 \nabla^2=\partial_{\rho}^2+4Q\partial_{\rho}, \qquad
 R=4(5Q^2+2\beta Q),
\eeq
where $Q(\lambda)$ is given by (\ref{QQ}). For the solutions (a) and (c)
in the previous section, $Q$ approaches to zero near $\lambda=0$. Then
we can solve the above equation (\ref{anom}) by the asymptotic form of
$\chi=e^{\Delta \rho}$ near $\rho\to \infty$ or $\lambda\to 0$
with 
\beq
  \Delta_{\pm}=\pm {1\over 2}\sqrt{-V_c}  .
\eeq
This implies $\beta(\lambda)=-\Delta_+\lambda+\cdots$. 

%===  Fig.4  ========================================================
\begin{figure}
\begin{center}
   \includegraphics[width=15cm,height=7cm]{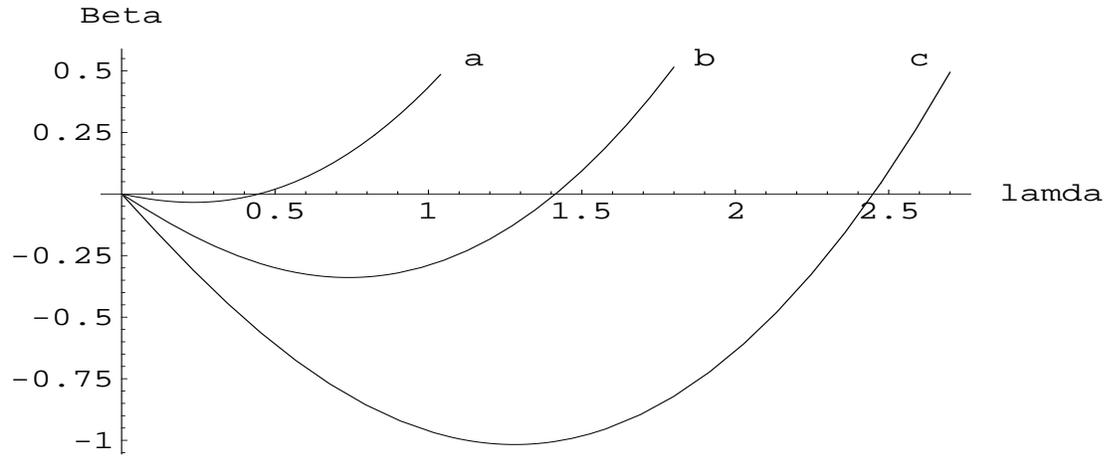}
 \caption{(a)-type solutions of $\beta(\lambda)$ obtained for 
a. $V_c=-.5/2$ b. $V_c=-5/2$ c. $V_c=-15/2$. The boundary conditions
are taken the same one with the solution (a) given in the Fig.1.
}
\end{center}
\end{figure}
%====================================================================
%===  Fig.5  ========================================================
\begin{figure}
\begin{center}
   \includegraphics[width=15cm,height=7cm]{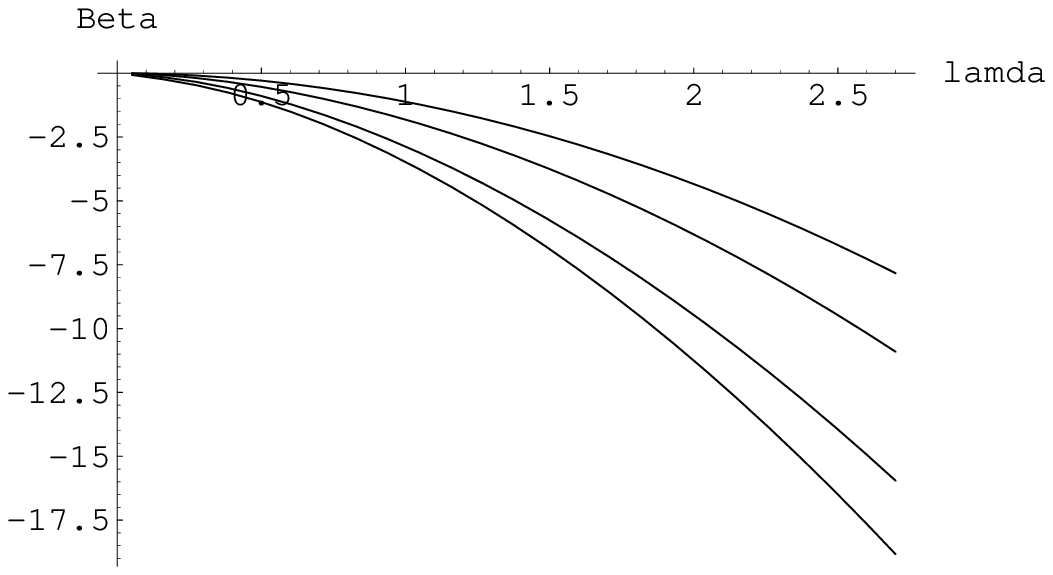}
 \caption{(b)-type solutions of $\beta(\lambda)$ obtained for 
a. $V_c=-5\times 10^{-4}/2$ b.$V_c=-.5/2$ c. $V_c=-5/2$ d. $V_c=-15/2$. 
The boundary conditions
are taken the same one with the solution (b) given in the Fig.1.
}
\end{center}
\end{figure}
%====================================================================
Then we must see the next order terms of $\beta(\lambda)$ to compare
with the perturbative behaviour of the Yang-Mills theory. We can obtain
the following form of the expansion by using Eq.(\ref{betafu2}),
\beq
  \beta(\lambda)=\beta_0^-\lambda
       +{\lambda^3\over 2\beta_0^-}\ln{\lambda\over\lambda_0}
       +{b_2\over 96(\beta_0^-)^3(2-\beta_0^-)}
           \lambda^5\left((\ln{\lambda\over\lambda_0})^2
         +a_2\ln{\lambda\over\lambda_0}+c_2\right)+\cdots \, , \label{series}
\eeq
\beq
  b_2=-21, \quad a_2=-{57\over 4}-{b_2\over 2(2-\beta_0^-)}, \quad
  c_2=-{9\over 128}-{a_2\over 2(2-\beta_0^-)},
\eeq
where $\lambda_0$ is an arbitrary constant. The second and the third terms
in (\ref{series}) represent the one-loop and the two-loop corrections 
respectively, and $b_2$ is not changed by the value of $\lambda_0$. The characteristic features of this result are the following two points. (i) This expansion
contains the logarithmic term, $\ln{\lambda\over\lambda_0}$, and (ii) the 
coefficient of the two-loop correction, $b_2$, is negative. The first point
implies the existence of the interaction term like 
$g_{\rm YM}^2A_{\mu}A_{\nu}O^{\mu\nu}$ 
in the gauge theory, and $O^{\mu\nu}$ condenses to the vacuum. Its value
would be related to $V_c$, because this behaviour disappears for $V_c=0$.
The similar
situation is seen in considering the Coleman-Weinberg mechanism for
a gauge theory.
However we can not say anything about the breaking of the gauge symmetry
in our case from this logarithmic behaviour only. 

The second point implies that the gauge theory considered here is not a
pure Yang-Mills theory with gauge fields only because of $b_2<0$. This
does not change the asymptotic freedom of the theory since the one-loop
coefficient is negative, but
the contents of the fields in the theory can not be cleared here.

The coefficients of the loop-corrections depend on the value
of $V_c$, and we can see this through $\beta$-functions of type (a) and (c)
given in the previous section by varing the values of $V_c$. The numerical
estimations for those are shown in the Figs.4 and 5. The qualitative
behaviors are not changed by the values of $V_c$ for (a) and (c) respectively.

Finally we comment on the identification of $\lambda$ with $g_{\rm YM}$.
It is given here as $\lambda\propto g_{\rm YM}$, which is different from the
case of $V_c=0$ where we obtain the expected correspondence
$\lambda\propto g_{\rm YM}^2$. Similar
mismatch is also seen in the case of the critical type 0B model, where
the relation $\lambda\propto g_{\rm YM}^4$ is preferred \cite{Min2,KT3}. 
Although the tachyon is treated as a running field in type 0B model, 
the value of $V_c$ is also negative and it plays an important role.
This kind of
ambiguity in identifying $\lambda$ and $g_{\rm YM}$ would be related
to the tachyon condensation which would lead to a deformation of the 
gauge coupling from the one expected by the naive world-volume
action of the D-branes.
%%%%%%%%%%%%%%%%%%%

%%%%%%%%%%%%%%%%%%%%%%%%%%%%%%%%%%%%%%%%%%%%%%%%%
\section{Stability of Tachyon}

Finally, we give a brief comment on the stability of the tachyon fluctuation
around our solutions.
Here the tachyon was not "running" since the classical equations
are solved for the case of $T=T_0$, which is independent on the 
energy scale $\rho$. But we must pay attention for 
its fluctuation, which is denoted by $t$, around $T_0$ from
the viewpoint of the stability of our classical solutions in
the five dimensional theory.
This is examined by solving the linearized 
equation for $t$, which is obtained from (\ref{tachyon2}) and
(\ref{dil}) as
\beq
    \ddot{t}+(d\dot{A}-2\dot{\phi})\dot{t}=
                                  {1\over 2}V''(T_0)t \, .
         \label{tachyonf}
\eeq
Here dot denotes the derivative with respect to the new variable $u$, but
it is equivalent to $\rho$ since we set as $du/d\rho=1$. The 
condition for the stability is the existence of the solution
of Eq.(\ref{tachyonf}) in the form $t=e^{\alpha u}$ with real 
$\alpha$. Since the coefficient of $\dot{t}$ is dependent on $u$, then we
examine this equation near the fixed points where the coefficient can be
approximated by a constant.

First, we consider near the asymptotic free region, $\lambda\sim 0$.
(i) For the solution of $V_c=0$, the coefficient disappears at $\lambda=0$,
namely $d\dot{A}-2\dot{\phi}=0$. Then $t$ will be stable if 
\beq
      V''(T_0)>0,  \label{positive}
\eeq
which means the positive mass-squared of the tachyon around $T=T_0$. 
(ii) For the solutions of $V_c<0$, solutions (a) and (c), 
we have $d\dot{A}-2\dot{\phi}=-2\beta_0^-$.
So the stability condition is obtained as
\beq
    {1\over 2}V''(T_0) -{V_c(T_0)\over 4} > 0 .  \label{condition}
\eeq
In this case, the condition is satisfied even if
$V''(T_0)$ is negative for its small absolute value. This situation
is similar to the case of the AdS background.

In the case of solution (a), the coefficient can be 
estimated at the AdS fixed point
$\lambda=\lambda_0$, where we get $d\dot{A}-2\dot{\phi}=3\lambda_0$. Then
the stability condition at this point is given by 
${1\over 2}V''(T_0) -{9V_c(T_0)\over 5} > 0$. Since $V_c<0$, this condition
is satisfied if (\ref{condition}) is fulfilled. Then we can say that the 
tachyon would be stable if the condition (\ref{condition}) is fulfilled
for the solution (a).

\section{Conclusions and Discussion}

We have examined
the equations of non-critical string theory
as the renormalization group equations of the Yang-Mills theory 
on the boundary.
The analysis is restricted to the five
dimensions to see the properties of the pure Yang-Mills
theory. The tachyon is assumed to be a constant to simplify the
model, and the equations are rewritten as the differential equations
with respect to the 't Hooft coupling constant $\lambda$.
In this equation, the value of the tachyon potential appears as
an unique parameter.

In terms of this simple model, we could find several interesting 
solutions.
The $\beta$-function of the corresponding Yang-Mills theory could have
two zero (or fixed) points 
at $\lambda=0$ and $\lambda_0$. The latter point ($\lambda_0$)
is corresponding to the fixed point with the AdS background. While,
the background at the
asymptotic-free fixed point ($\lambda=0$) is the flat space or non-AdS
space with zero curvature.
Among the various solutions, two types (called as type (a) and (c))
of asymptotic-free solutions are found. They are classified
by their infrared behaviour. For the solution of type (c), 
the $\beta$-function decreases
monotonically with increasing $\lambda$. While the type (a) solution connects
two fixed points at $\lambda=0$ and $\lambda_0$ with different background 
configurations smoothly, and the
asymptotic-free phase is restricted to the finite region of $\lambda$.

For the type (a) and the analytic type (c) solutions,
the ZigZag horizons have been found at the infrared and ultraviolet
fixed point respectively. 
The Wilson loops have been estimated for these solutions
and we found the quark-confinement potentials for both solutions. 
But the potential obtained grows with the distance between
quarks more slowly than the linear rising one, and 
the Coulomb like potential at small distance could not 
be seen for the latter case.

%%%%%%%%
For the analytic solution of $V_c=0$, we can see the expected scale
dependence of the gauge coupling constant, $g_{\rm YM}$, and 
its identification with the 'tHooft coupling as
$\lambda=g_{\rm YM}^2$. However we observe several unexpected features
of the solutions for $V_c\neq 0$. They can not be seen in the
usual pure Yang-Mills theory. (i) The leading term of $\beta(\lambda)$
near $\lambda=0$ is the linear term of $\lambda$, and it would represent
the anomalous dimension of the coupling constant. This can be
understood by the wave equation of the dilaton, $e^{\Phi}$, near
the boundary of the bulk space. In the background of our solution,
the dilaton behaves as a massive field. The mass has been produced
effectively by the background configuration, and the mass term produces an
anomalous dimension for $e^{\Phi}$. (ii) The loop-expansions
contains the logarithmic factor of the coupling constant. This fact
implies that the gauge theory considering here would contain some operator
which couples to the gauge fields and condenses to the vacuum as seen
in the Coleman-Weinberg mechanism. (iii) The 'tHooft coupling should be
identified with the gauge coupling constant as 
$\lambda\propto g_{\rm YM}$. This identification is different from the case
of the solution for $V_c=0$. Then we expect that the condensation of the
tachyon would lead to some modification of the gauge coupling constant.
%%%%%%%%%%%%%%%%

The availability of our analysis based on the holographic idea would be 
restricted to the small $\lambda$ region as mentioned in section four.
In this sense, the solution of type (a) 
would be reliable even in the infrared region if we consider
in the asymptotic free phase and $V_c$ is taken to be small.

We could obtain the stability condition for the tachyon-fluctuation
around our solutions
at the ultraviolet and infrared fixed points, 
and we found that condition would be satisfied even if tachyon has 
negative mass-squared as in the AdS background case.

\vspace{.5cm}

%\noindent {\bf Acknowledgement}

%\vspace{.3cm}
%  The author thanks for their useful comments.

\end{document}